\documentclass[10pt,letterpaper]{article}
\usepackage{opex3}

\usepackage{amsmath}
\usepackage{cite}

\begin{document}
\title{Emission of photon echoes in a strongly scattering medium}
\author{F. Beaudoux,$^1$ A. Ferrier,$^1$
O. Guillot-No\"el,$^{1,3}$ T. Chaneli\`ere,$^2$\\ J.-L. Le Gou\"et,$^2$
 and Ph. Goldner$^{1,\ast}$} 
\address{$^1$Chimie-Paristech, 
  Laboratoire de Chimie de la Mati\`ere Condens\'ee de Paris, CNRS-UMR
  7574, UPMC Univ Paris 06, 11 rue Pierre et Marie Curie 75005 Paris, France}
\address{$^2$Laboratoire Aim\'e Cotton, CNRS-UPR 3321,
  Univ. Paris-Sud, B\^at. 505, 91405 Orsay cedex, France}
\address{$^3$in memoriam}
\email{$^\ast$philippe-goldner@chimie-paristech.fr}

\begin{abstract}
  We observe the two- and three-pulse photon echo emission from a
  scattering powder, obtained by grinding a Pr$^{3+}$:Y$_2$SiO$_5$
  rare earth doped single crystal. We show that the
  collective emission is coherently constructed over several grains. A
  well defined atomic coherence can therefore be created between
  randomly placed particles. Observation of photon echo on powders as
  opposed to bulk materials opens the way to faster material
  development. More generally, time-domain resonant four-wave mixing
 offers an attractive approach to investigate
    coherent propagation in scattering media.
\end{abstract}


\bibliographystyle{osajnl}
\bibliography{PowdersPr_bib,quantum_info}

\section{Introduction}
\label{sec:introduction}

Photon echo~\cite{Kurnit1964,Abella1966} refers to a time-delayed non-linear coherent optical response to resonant excitation by a specific sequence of light pulses. At first sight, the coherent buildup of such a signal should require high quality optical materials. However, optical waves can propagate coherently in random media. This has attracted considerable attention for decades and is revealed by features such as the backscattering peak~\cite{albada1985,Wolf1985} or the random laser~\cite[and references therein]{wiersma2008physics} in the multiple scattering regime. Coherent propagation in random media can also be combined with non-linear optical
processes~\cite{kravtsov1991theory,baudrier2004random}. The observation of localization effects, whether reduced, enhanced or simply tested by non-linear processes, definitely opens new perspectives~\cite{deBoer1993,vanneste2001selective,cao2000spatial,wellens2008,wellens2009}. In the specific framework of four-wave mixing (FWM), the class of non-linear processes to which photon echo belongs \cite{mossberg1982time}, it has been recognized quite early that coherent anti-Stokes Raman scattering (CARS) can be observed in polycrystalline and opaque media~\cite{Dlott1991}. There is considerable practical interest in accomodating coherent optics to disordered materials. Since many interesting chemical solids are difficult to crystallize, substituting a rough powder to a highquality monocrystal would expedite the testing of new compounds~\cite{markushev1990stimulated}. It may also prove useful to probe raw materials with sophisticated techniques. Quite recently, CARS has been proposed for fast detection of bacterial endospores in the context of bacterial warfare~\cite{Zhi2007}.

In the continuation of CARS investigations, we report on the first observation, to the best of our knowledge, of photon echo emission from a polycrystalline opaque powder. This offers an attractive access to the optical dipole lifetime $T_2$ in cheap and easily produced new compounds in the prospect of classical and quantum processing \cite{colice2004rf,LeGouet2007,tittel-photon,riedmatten2008}. Moreover, we consider the contribution of photon echo features, namely the time ordering and the frequency degeneracy, to the study of coherent collective emission in disordered materials. Rather than investigating the scattering properties of the incoming fields, we analyze the matching of the echo mode with the scattered fields. In a similar way, spatial correlations have been extensively exploited to understand light propagation in random media~\cite{Feng1988,ito2004}.

\section{Experiment}
\label{sec:experiment}

The photon echo is generated on the $^3$H$_4$-$^1$D$_2$ transition of Pr$^{3+}$:Y$_2$SiO$_5$, cooled down to liquid helium temperature. A first pulse resonantly excites Pr$^{3+}$ ions at time $t_{I}$. Since the transition is inhomogeneously broadened by the crystal field, the excited dipoles  oscillate at different frequencies and dephase from each other. A second pulse is shined to the sample at time $t_{II}$. This reverses the dipole phase-shift, regenerates the macroscopic polarization and gives rise to a restored radiative response at time $2t_{II}-t_{I}$. As a function of the time interval $t_{II}-t_{I}$, the delayed signal decays with time constant $T_2$, corresponding to the optical dipole lifetime. Access to $T_2$ represents the main interest of photon echo.

\begin{figure}
  \centering
 \includegraphics[width=7cm]
{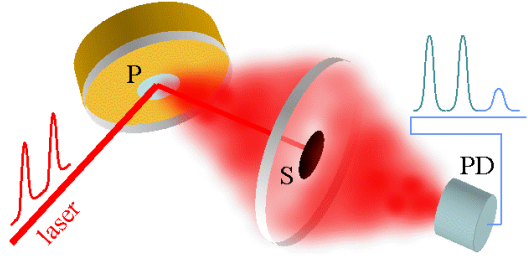}
  \caption{Experimental set-up. P: powder, S: light stop, PD: photodetector.}
  \label{fig:setup}
\end{figure}

In the present experiment, the usual single crystal is substituted with a powder obtained by grinding and sieving a 0.05 at.\% Pr:Y$_2$SiO$_5$ single crystal. The average grain size is 55($\pm15$)$\mu$m. The experimental set-up is shown in Fig. \ref{fig:setup}. The powder fills a 2 mm diameter, 0.5 mm depth hole in a copper holder and is maintained by two glass plates. The holder is placed into a bath cryostat and kept at 3.1 K. Excitation of site 1 at 605.97 nm (vacuum) is provided by a 1 MHz linewidth Coherent 899-21 dye laser. Pulse shaping and frequency shifting are provided by an acousto-optic modulator mounted in double pass configuration. With the help of a 125 mm focal length lens, the 2 mm diameter beam is focused on the powder to a spot size of $\approx$ 50 $\mu$m (gaussian beam diameter at 1/e$^{2}$ the axial intensity). The powder surface is set at 45$^\circ$ from the beam propagation direction in order to direct part of the scattered light through a side window of the cryostat. The scattered light is then collected and focused onto an avalanche photodiode with a 10 MHz bandwidth. An opaque screen is used to stop the laser light directly reflected by the front glass plate. The light is strongly scattered by the powder that only transmits about 10$^{-4}$ of the incident laser intensity. The corresponding mean free path is 55 $\mu$m with is consistent with the average grain size. The scattered light shows a speckle pattern in the far field, as expected in the coherent propagation regime.

\section{Optical dipole lifetime measurement}
\label{sec:results-discussion}

\begin{figure}
  \centering
 \includegraphics
[width=8cm]{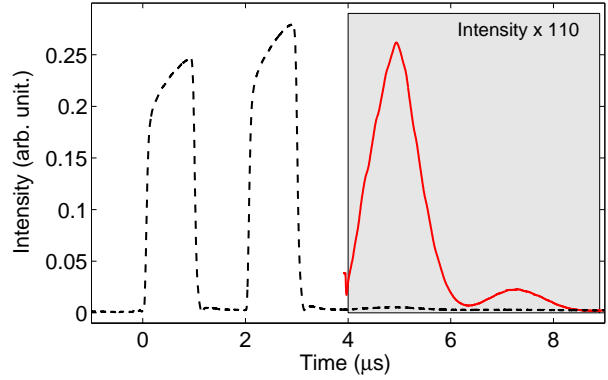}
  \caption{Scattered light intensity recorded for a two-pulse photon echo sequence in a Pr$^{3+}$:Y$_2$SiO$_5$ powder. Dashed line: Excitation pulses. Solid line: two-pulse echo (intensity $\times$ 110). The incident laser power is $32$ mW.}
  \label{fig:twopulsePE}
\end{figure}

The scattered light is recorded when two identical 1 $\mu$s long square pulses, separated by 1 $\mu$s, are shined to the powder (Fig. \ref{fig:twopulsePE}). During the pulses, the scattered intensity grows, which reflects the saturation of the optical transition (bleaching).
To avoid hole burning effects due to population trapping in Pr$^{3+}$ ground state hyperfine levels, we scan the laser frequency over 1GHz in about 2 s. At the expected temporal position of the photon echo, between $t=4$ and $t=6$ $\mu$s, we observe a signal that disappears when the first pulse is suppressed or when the laser is detuned from the absorption band. When the pulse separation is increased, the signal is delayed accordingly, appearing at the expected $2t_{II}-t_{I}$ time. These features ascertain the photon echo nature of the signal. About $2$ $\mu$s after the first one, a secondary echo can be observed, the first echo acting as an excitation pulse.

In order to determine $T_2$, we record the echo decay as a function of the time separation between the driving pulses.  At 3.1 K, the
exponential decay leads to a coherence lifetime of 53 $\mu$s.
This value is smaller than in the bulk crystal, where we measured $T_2= 80$ $\mu$s. The strains induced by grinding at the particle level might explain this discrepancy. Indeed, the inhomogeneous linewidth increases from 5.5 in the bulk, to 9 GHz in the powder. The strains could create low frequency tunneling modes which in turn are known to induce dephasing, even at low temperatures \cite{Macfarlane2004}. Surface related defects or pollutions are unlikely to be significant, given the relatively large size of the grains.

\section{Coherent build-up}\label{Coherent_build-up}
At this stage we ignore in which spatial mode the echo is radiated. The different grains may contribute incoherently to the detected signal. Then, buildup is spatially coherent at the scale of each individual particle only. The different grains give rise to independent signals that are incoherently scattered to the sample output. Instead, coherence may be preserved all along the propagation path through the powder. In this alternative picture, the echo builds up coherently along a distance significantly larger than the grain size. By heterodyning the echo with a reference field we are able to decide which picture is valid.

Heterodyning is performed in the general three-pulse echo (3PE) configuration. The two-pulse echo (2PE) is a special case in which the second and third pulses coincide. The sample is successively shone by three excitation pulses, labeled $I$, $II$, $III$ according to their time ordering. They are directed to the sample along wavevectors $\textbf{k}_{I}$, $\textbf{k}_{II}$ and $\textbf{k}_{III}$ respectively, with $k_{I}=k_{II}=k_{III}=k$ and reach the target at times $t_{I}$, $t_{II}$ and $t_{III}$, respectively. The echo corresponds to the diffraction of the third pulse on the spectro-spatial grating that has been engraved in the material by the first two pulses. The signal emission occurs at $t_{III}+t_{II}-t_{I}$ when all the excited dipoles are phased together. 

We set two beams $\textbf{k}_{1}$ and $\textbf{k}_{2}$ from the laser, at angle $\theta=80$ mrad. With an acousto-optic switch on each beam we are able to shape any desired pulse sequence. All pulses are rectangular and 1 $\mu$s long. They are separated by the time intervals $t_{II}-t_{I}=1\mu$s and $t_{III}-t_{II}=10\mu$s. 

When pulses $I$ and $III$ impinge the powder along $\textbf{k}_{1}$, while pulse $II$ is directed along $\textbf{k}_{2}$, we observe an echo at time at $t_{III}+t_{II}-t_{I}$. In non-scattering medium, the echo should be emitted with the well defined wavevector $\textbf{k}_e=\textbf{k}_{III}+\textbf{k}_{II}-\textbf{k}_{I}$, which equals $\textbf{k}_{2}$ when pulses $I$ and $III$ copropagate. In the present situation the signal is just scattered over a broad angle. We heterodyne the echo with a weak probe pulse that we shine to the powder along $\textbf{k}_{2}$, during the signal duration. This is reminiscent of speckle correlation experiments~\cite{Feng1988,ito2004}, except that we presently correlate fields instead of intensities. 

\begin{figure}
  \centering
  \includegraphics[width=8cm]{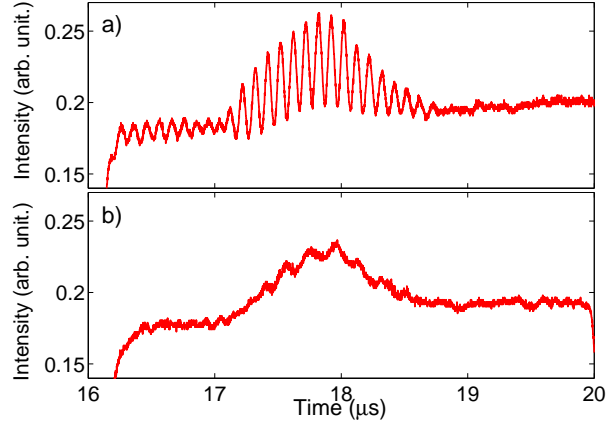}
  \caption{Three-pulse photon echo heterodyne detection in situation (i): $\textbf{k}_{I}=\textbf{k}_{III}=\textbf{k}_{1}$ and $\textbf{k}_{II}=\textbf{k}_{2}$. A 10 MHz-shifted probe pulse is used as a local oscillator (LO). (a) Beating is observed when the LO is sent along $\textbf{k}_{II}$. (b) No beating is observed with a LO along $\textbf{k}_{I}$.}
  \label{fig:het}
\end{figure}

As illustrated in Fig. \ref{fig:het}.a, a strong beating is observed. We also verify the absence of beating with the other beam $\textbf{k}_{I}$ (Fig. \ref{fig:het}.b). This clearly means that the echo field and the scattered pulse $II$ spatial modes coincide at the powder output. In other words, the echo coherently builds up through the material, all along the path followed by pulse $II$. It is known that heterodyne detection is very demanding in terms of spatial mode identity~\cite{Gorju2005} and one cannot help but feel surprised by the high contrast of the interference pattern. Indeed, the scattered light is collected over an angular aperture $\Theta$ of several hundreds milliradians, which corresponds to a spatial resolution of a few wavelengths at the sample surface. In other words, the spatial modes are compared down to the finest detail level, where they remarkably coincide.

In the general context of non-linear optics in scattering media, it should be stressed that the echo selective heterodyning is only made possible by the temporal separation of the successive pulses. The non-linear signal is detected on dark background, after the extinction of the driving fields. 

\section{Phase matching}
In bulk non-scattering medium, the echo wavevector $\textbf{k}_e=\textbf{k}_{III}+\textbf{k}_{II}-\textbf{k}_{I}$ must satisfy the phase matching condition $\left|k_e-k\right|L<\pi$, where $L$ stands for the sample optical length. The requirement is fulfilled in situation $(i)$, when $\textbf{k}_{I}$ coincides with either $\textbf{k}_{II}$ or $\textbf{k}_{III}$ and the signal intensity does not vary with angle $\left(\textbf{k}_{II},\textbf{k}_{III}\right)$. On the contrary, in situation $(ii)$, when $\textbf{k}_{II}=\textbf{k}_{III}$, the signal drops to zero as soon as $\left(\textbf{k}_{I},\textbf{k}_{II}\right)$ exceeds $\sqrt{\pi/(kL)}$.

\begin{figure}
  \centering
  \includegraphics
[width=10cm]{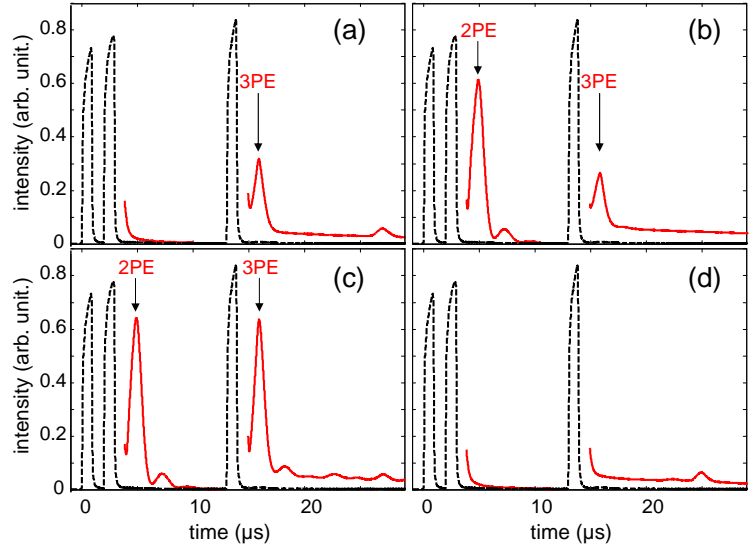}
\caption{Scattered light intensity recorded for a three-pulse photon echo (3PE) sequence for different excitation pulses time-ordering. The dashed line represents the excitation pulses. Solid lines: (a) situation (i): $\textbf{k}_{I}=\textbf{k}_{III}=\textbf{k}_{1}$ and $\textbf{k}_{II}=\textbf{k}_{2}$; (b) $\textbf{k}_{I}=\textbf{k}_{II}=\textbf{k}_{1}$ and $\textbf{k}_{III}=\textbf{k}_{2}$; (c) $\textbf{k}_{I}=\textbf{k}_{II}=\textbf{k}_{III}=\textbf{k}_{1}$. The phase matching condition is satisfied for (a), (b) and (c). 
 Two-pulse echos (2PE) are present in cases (b) and (c) (see text for details). (d) Situation (ii): 3PE is not observed.}
  \label{fig:threePE}
\end{figure}

In the angled beam configuration described above, we observe similar features in the scattering medium, as shown in Fig. \ref{fig:threePE}. The echo is detected in situation $(i)$, when $\textbf{k}_{I}=\textbf{k}_{III}=\textbf{k}_{1}$ and $\textbf{k}_{II}=\textbf{k}_{2}$ (Fig. \ref{fig:threePE}.a), or when  $\textbf{k}_{I}=\textbf{k}_{II}=\textbf{k}_{1}$ and $\textbf{k}_{III}=\textbf{k}_{2}$ (Fig. \ref{fig:threePE}.b), or in the degenerate case  $\textbf{k}_{I}=\textbf{k}_{II}=\textbf{k}_{III}=\textbf{k}_{1}$ (fig. \ref{fig:threePE}.c). It disappears in situation $(ii)$ when $\textbf{k}_{I}=\textbf{k}_{2}$ and $\textbf{k}_{II}=\textbf{k}_{III}=\textbf{k}_{1}$(Fig. \ref{fig:threePE}.d). 

\begin{figure}
  \centering
  \includegraphics
[width=8cm]{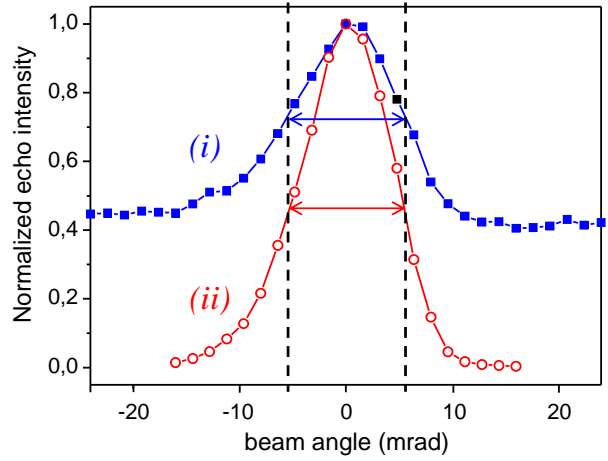}
\caption{Echo intensity variations with beam angle $\theta=\left(\textbf{k}_1,\textbf{k}_2\right)$. In situation $(i)$, where $\textbf{k}_{I}=\textbf{k}_{III}=\textbf{k}_{1}$ and $\textbf{k}_{II}=\textbf{k}_{2}$, the signal decreases to half the maximum value at large angle. In situation $(ii)$, where $\textbf{k}_{I}=\textbf{k}_{2}$ and $\textbf{k}_{II}=\textbf{k}_{III}=\textbf{k}_{1}$, the signal drops to zero at large angle. The characteristic angle of variation exhibits the same size in both profiles.}
  \label{fig:correlation}
\end{figure}

We further investigate the echo intensity variations with the $\theta=\left(\textbf{k}_1,\textbf{k}_2\right)$ beam angle. The results are displayed in Fig. \ref{fig:correlation}. In situation $(ii)$, the signal drops to 0 when the beam angle is increased, as expected. Unexpectedly, the echo intensity also depends on $\theta$ in situation $(i)$, reaching a steady level of about half maximum at large angle. In both cases the echo decreases over an angular range of $\approx5$ mrad. The one-half reduction in situation $(i)$ is robust against many changes in the experimental conditions. For instance this ratio is preserved when driving pulse intensity is varied over one decade, from weak field to saturation regime. The same ratio is also observed when the echo is observed by transmission through the powder instead of scattering from the surface. We verified that the echo intensity at $\theta=0$ does not depend whether the incoming pulses are shaped from a single or from two different beams. This rules out a possible spatial mode mismatch of the driving fields. 

Within the framework of the bulk material picture, the 5 mrad half-width should be identified with the cut-off angle $\sqrt{\pi/(kL)}$, leading to a propagation distance $L$ larger than 10 mm. This length can be converted into an optical depth $\alpha L > 11$, given the 11 cm$^{-1}$ average value of the $\alpha$ absorption coefficient in the bulk crystal used to obtain the powder. Such a large optical depth strongly disagrees with a more direct measurement based on the saturation of the absorbing transition. Only partially bleached by the driving pulses in Fig.\ref{fig:twopulsePE}, the material can be made totally transparent by a long enough pulse, able to equalize the population in the ground and excited levels. Hence, along the pulse duration, the scattered intensity grows by the factor $\mathrm{e}^{\alpha L}$, which gives access to the optical depth. This way we obtain $\alpha L=0.23$, about 50 times smaller than the value derived from the cut-off angle. This shows the inadequacy of the bulk material picture. As soon as the beams enter the powder, scattering strongly modifies their spatial mode. Any correlation between them is probably lost as soon as their relative angle of incidence exceeds their diffraction limited aperture. The latter quantity, of order of the wavelength to the laser-spot diameter ratio, is actually close to the characteristic angle of echo variations.
     
The discrepancies of experimental data with the bulk description invite to develop a specific analysis.  
\section{Tentative description of the echo build-up}
In the weak field limit, the contribution at position $\textbf{r}$ to the echo signal is proportional to:
\[ 
E_I^*(\textbf{r})E_{II}(\textbf{r})E_{III}(\textbf{r})
\]
where $E_j(\textbf{r})$ denotes the spatial distribution of driving field $\#j$. Let us first consider the degenerate situation where the three driving fields impinge the powder in the same spatial mode $E_I(\textbf{r})=E_{II}(\textbf{r})=E_{III}(\textbf{r})$, traveling along the same wavevector. Then either $E_I^*(\textbf{r})E_{II}(\textbf{r})$ or $E_I^*(\textbf{r})E_{III}(\textbf{r})$ can be identified with $I(\textbf{r})$, the driving intensity at position $\textbf{r}$. Because of scattering, $I(\textbf{r})$ exhibits a speckle structure, with a correlation length of about one wavelength. However, as demonstrated in Section~\ref{Coherent_build-up}, the signal builds up coherently, propagating in the same mode as one of the driving fields. Therefore each local contribution must keep the spatial dependence of a single field. This can only proceed from averaging $I(\textbf{r})$ over the light spot. Thus the local contribution to the coherent part of the echo reads as:
\begin{align}\label{local_echo} 
\left[E_I^*(\textbf{r})E_{II}(\textbf{r})E_{III}(\textbf{r})\right]_\mathrm{coh.}&=\frac{1}{2}<E_I^*(\textbf{r})E_{II}(\textbf{r})>E_{III}(\textbf{r})+\frac{1}{2}<E_I^*(\textbf{r})E_{III}(\textbf{r})>E_{II}(\textbf{r})\\
&=\frac{1}{2}<I>\left[E_{II}(\textbf{r})+E_{III}(\textbf{r})\right]
\end{align}
where average is denoted by $<>$. This way one neglects the incoherent contribution, proportional to $\delta I(\textbf{r})=I(\textbf{r})-<I>$, which represents the $\textbf{r}$-dependent fluctuations of $I(\textbf{r})$.  

Next, let us consider the angled-beam situation $(i)$, assuming $\textbf{k}_{I}=\textbf{k}_{III}=\textbf{k}_{1}$ and $\textbf{k}_{II}=\textbf{k}_{2}$. Hence $E_I^*(\textbf{r})E_{II}(\textbf{r})$ no longer coincides with $E_I^*(\textbf{r})E_{III}(\textbf{r})$. Instead, when $\theta=\left(\textbf{k}_1,\textbf{k}_2\right)$ exceeds the diffraction limited aperture of the incoming beams, $<E_I^*(\textbf{r})E_{II}(\textbf{r})>$ vanishes. Therefore the local coherent contribution in Eq.~\ref{local_echo} reduces to $(1/2)<E_I^*(\textbf{r})E_{III}(\textbf{r})>E_{II}(\textbf{r})=(1/2)<I>E_{II}(\textbf{r})$, which corresponds to half the $\theta=0$ value. This represents only 1/4 the $\theta=0$ intensity, two times smaller than the experimental observation. 

Finally, in situation $(ii)$, when $\textbf{k}_{II}=\textbf{k}_{III}=\textbf{k}_{1}$ and $\textbf{k}_{I}=\textbf{k}_{2}$, both $E_I^*(\textbf{r})E_{II}(\textbf{r})$ and $E_I^*(\textbf{r})E_{III}(\textbf{r})$ vanish on average when $\theta$ exceeds the diffraction limited aperture, which completely cancels the local coherent contribution. The echo vanishes, as observed experimentally. In this picture, the signal variation with $\theta$ reflects spatial mode correlation features rather than phase matching properties. 

The incoming spatial mode is assumed to be gaussian. In terms of the $d=2$ mm laser beam diameter at the focusing lens with $f=12.5$ cm focal length, the full width at half maximum of the echo variations with $\theta$ can be expressed as $\sqrt{\mathrm{Ln}(2)/2}\;d/f\approx9$ mrad, in excellent agreement with experimental data.   

The analysis correctly describes some features but fails to provide the robust 1/2 signal intensity reduction we observe in situation $(i)$ when we increase the beam angle.

\section{Conclusion}
\label{sec:conclusion}

In conclusion, we have observed photon echoes in a scattering crystalline powder. This opens the way to testing new rare earth doped materials in the shape of crystalline powders, considerably easier and faster to synthesize than single crystals. This technique could also be applied to systems, such as molecules, that one is usually forced to disperse in a solvant to get a non scattering solution. By experimenting directly in powders, one could determine and/or suppress the dephasing effect of the solvant. Finally, by determining the echo spatial mode at the sample output, we have proved that the signal coherently builds up all along propagation through the powder. Further theoretical investigation is needed to clarify the signal behavior in angled-beam congiguration.  

\section{Acknowledgements}
\label{sec:acknowledgements}

This work is supported by the European Commission through the FP7 QuRep project.

\end{document}